\pdfoutput=1
\documentclass[12pt,letterpaper,floatfix]{article}
\usepackage{amsmath}
\usepackage{amssymb}
\usepackage{epsfig}
\usepackage{euscript}
\def\mathswitchr#1{\relax\ifmmode{\mathrm{#1}}\else$\mathrm{#1}$\fi}

\newcommand {\pslash}{\hbox{$\not\hbox{\kern-2.3pt $p$}$}}

\newcommand{\FYFS}{F_{\mathrm{YFS}}}
\usepackage{cite}

\usepackage{epic}

\def\alf1{ {\alpha\over\pi} }

\begin{document}
\begin{titlepage}
\begin{flushleft}
{\bf BU-HEPP-15-01}\\
{\bf Mar., 2015}\\
\end{flushleft}
\begin{center}
{\bf \large IR-Improved DGLAP-CS QCD Parton Showers in Pythia8}\\
\vspace{2mm}
B.F.L. Ward\\%
      Baylor University, Waco, TX, USA\\
        E-mail: bfl\_ward@baylor.edu\\
\end{center}
\vspace{2mm}
\centerline{\bf Abstract}
We introduce the recently developed IR-improved DGLAP-CS theory into the 
showers in Pythia8, as this Monte Carlo event generator is in wide use at 
LHC. We show that, just as it was true in the IR-improved shower Monte Carlo 
Herwiri, which realizes the IR-improved DGLAP-CS theory in the Herwig6.5 
environment, the soft limit in processes such as single heavy gauge boson 
production is now more physical in the IR-improved DGLAP-CS theory
version of Pythia8. This opens 
the way to one's getting a comparison between the actual detector simulations 
for some of the LHC experiments between IR-improved and unimproved showers 
as Pythia8 is used in detector simulations at LHC whereas Herwig6.5, the 
environment of the only other IR-improved DGLAP-CS QCD MC in the literature, 
Herwiri1.031, is not any longer so used. Our achieving the availability of 
the IR-improved DGLAP-CS 
Pythia8 then is an important step in the further development 
of the LHC precision theory program under development by the author and his 
collaborators.  \\
\vspace{1cm}
%
\end{titlepage}
%

 
\def\Kmax{K_{\rm max}}\def\ieps{{i\epsilon}}\def\rQCD{{\rm QCD}}


In a series of papers~\cite{herwiri1,herwiri2,herwiri3,herwiri4}, 
we and our collaborators have developed, implemented and applied to FNAL and LHC data 
the IR-improved~\cite{irdglap1,irdglap2} DGLAP-CS~\cite{dglap,cs} theory 
in the 
Herwig6.5~\cite{hrwg} environment as realized
in the new Monte Carlo Herwiri1.031. Because the IR-improvement in Herwiri1.031 derives from the exact amplitude-based resummation theory in Refs.~\cite{qced}\footnote{The reader interested in the chronology of the theory can see Refs.~\cite{yfs,yfs-jw} for the original Abelian gauge theory development and application of the approach; here, the non-Abelian generalization is needed.} we and our collaborators have argued~\cite{herwiri1,herwiri2,herwiri3,herwiri4} that Herwiri1.031 
should and does give a better fit to the FNAL and LHC data on single heavy gauge boson production without the need of an ad hocly hard intrinsic $p_T$ spectrum (rms $p_T\simeq 2$ GeV/c) for the proton or anti-proton constituents, as one expects from observations like the precociousness of Bjorken scaling~\cite{scaling,bj1}. As we and our collaborators continue with the comparisons between Herwiri1.031 predictions and the recent LHCb data~\cite{1412-8717} on single heavy boson production and decay, we have met a matter of some concern as follows.\par
In some of the spectra which depend on the transverse degrees of freedom of the 
heavy gauge bosons, detector related effects such as 
bin migration are based on the 
detector simulations with the events of only some specific MC's and there is 
considerable over-head to re-do such simulations with Herwiri1.031 events 
because it uses the Herwig6.5 environment whereas these detector effect 
modules do not use that environment currently. Thus, it is somewhat ill-timed 
to get IR-improved showers via Herwiri1.031 into the LHC detector simulations 
for such effects as these important bin migration effects. We stress that, 
since the MC's for the IR-improved and unimproved showers look very different 
in the soft regime where these migration effects tend to be more pronounced, 
it is important to provide a platform which will facilitate the comparison 
between IR-improved and unimproved DGLAP-CS showers in this regard.\par
Accordingly, we have undertaken\footnote{We thank here Dr. Jesper Christiansen 
for useful private communications.} the introduction of the 
IR-improved DGLAP-CS theory into the Pythia8~\cite{pythia8} environment which 
, at least currently, is more widely used in 
detector simulation studies at LHC. In 
this Letter, we describe the introduction and illustrate its effect on the 
proto-typical heavy single $Z/\gamma^*$ production $p_T$ spectrum at the LHC. 
The detailed phenomenological studies will appear elsewhere~\cite{elswh}.\par
Specifically, the IR-improved DGLAP-CS theory is given in detail in Refs.~\cite{herwiri1,herwiri2,herwiri3,herwiri4,irdglap1,irdglap2}, so that we will not repeat it here and we refer the reader to the latter references for its specification. We turn directly to what is needed to introduce the theory into
the showers in Pythia8.\par
Toward this end, we proceed as follows. Focusing first on the time-like showers in Pythia8, 
in the module {\rm TimeShower.cc} we replace the usual DGLAP-CS kernels
with the IR-improved ones in Eqs.(6) in Ref.~\cite{herwiri2}. For example, 
whenever we have the shower weight factor $(1+z^2)$ (note that $z$ here is
{\rm dip.z} in {\rm TimeShower.cc}) for a given type of QCD
radiator color representation $A$ with the attendant infrared point at $z\rightarrow 1$, we make the replacement
\begin{equation}
(1+z^2) \rightarrow \FYFS(\gamma_A)e^{\frac{1}{2}\delta_A}(1+z^2)(1-z)^{\gamma_A},
\label{eq-pyiri-1}
\end{equation}
where the IR improvement exponents $\gamma_A,\delta_A$ and the YFS infrared function $\FYFS(x)$ are given in Eqs.(7) and (8) in Ref.~\cite{herwiri2}.\par
Continuing in this way, when we meet the shower weight factor $(1+z^3)$ in the gluon, $G$, splitting to $G\;G$ with the infrared point at $z\rightarrow 1$, we make the replacement
\begin{equation}
(1+z^3) \rightarrow \FYFS(\gamma_G)e^{\frac{1}{2}\delta_G}(1+z^3)(1-z)^{\gamma_G}.
\label{eq-pyiri-2}
\end{equation}
Finally, when we meet the shower weight factor $(z^2+(1-z)^2)$ in the splitting $G\rightarrow q \bar{q}$ we  make the replacement
\begin{equation}
(z^2+(1-z)^2)\rightarrow \FYFS(\gamma_G)e^{\frac{1}{2}\delta_G}(z^2(1-z)^{\gamma_G}+(1-z)^2z^{\gamma_G}).
\label{eq-pyiri-3}
\end{equation}
These changes in module {\rm TimeShower.cc} realize the IR-improved DGLAP-CS theory in time-like showers in Pythia8.\par
Turning next to the space-like showers in Pythia8, we act on the module
{\rm SpaceShower.cc} as follows. When we meet the shower weight factor $(1-z(1-z))^2$
in the splitting $G\rightarrow G\;G$, we make the replacement{\small
\begin{equation}
(1-z(1-z))^2\rightarrow \FYFS(\gamma_G)e^{\frac{1}{2}\delta_G}\left((1-z)^2z^{\gamma_G}+z^2(1-z)^{\gamma_G}+\frac{1}{2}z^2(1-z)^2(z^{\gamma_G}+(1-z)^{\gamma_G})\right).
\label{eq-pyiri-4}
\end{equation}}
When we meet the shower weight factor $(1+(1-z)^2)$ in the splitting $q\rightarrow G(z)\; q$ we make the replacement
\begin{equation}
\left(1+(1-z)^2\right)\rightarrow \FYFS(\gamma_q)e^{\frac{1}{2}\delta_q}\left(1+(1-z)^2\right)z^{\gamma_q}.
\label{eq-pyiri-5}
\end{equation} 
Continuing in this way, when we meet the shower weight factor $(1+z^2)$ in the splitting $q\rightarrow q(z) \; G$ we make the replacement
\begin{equation}
\left(1+z^2\right)\rightarrow \FYFS(\gamma_q)e^{\frac{1}{2}\delta_q}\left(1+z^2\right)(1-z)^{\gamma_q}.
\label{eq-pyiri-6}
\end{equation}
We note as well that mass corrections for the heavier quarks also receive the same IR improvement factors here. For the splitting $G\rightarrow q \; \bar{q}$ we make the replacement of the splitting weight factor $(z^2+(1-z)^2)$ as indicated above in (\ref{eq-pyiri-3}) for light quarks. For massive quarks, the corresponding mass correction in the weight factor has its factor of $2z(1-z)$ replaced according to the rule:
\begin{equation}
\left(2z(1-z)\right)\rightarrow \FYFS(\gamma_G)e^{\frac{1}{2}\delta_G}\left(z(1-z)(z^{\gamma_G}+(1-z)^{\gamma_G})\right).
\label{eq-pyiri-7}
\end{equation}   
With these replacements in module {\rm SpaceShower.cc}, we have introduced the IR-improved DGLAP-CS theory into the space-like
showers of Pythia8.\par
\begin{figure}[h]
\begin{center}
\includegraphics[width=100mm,angle=90]{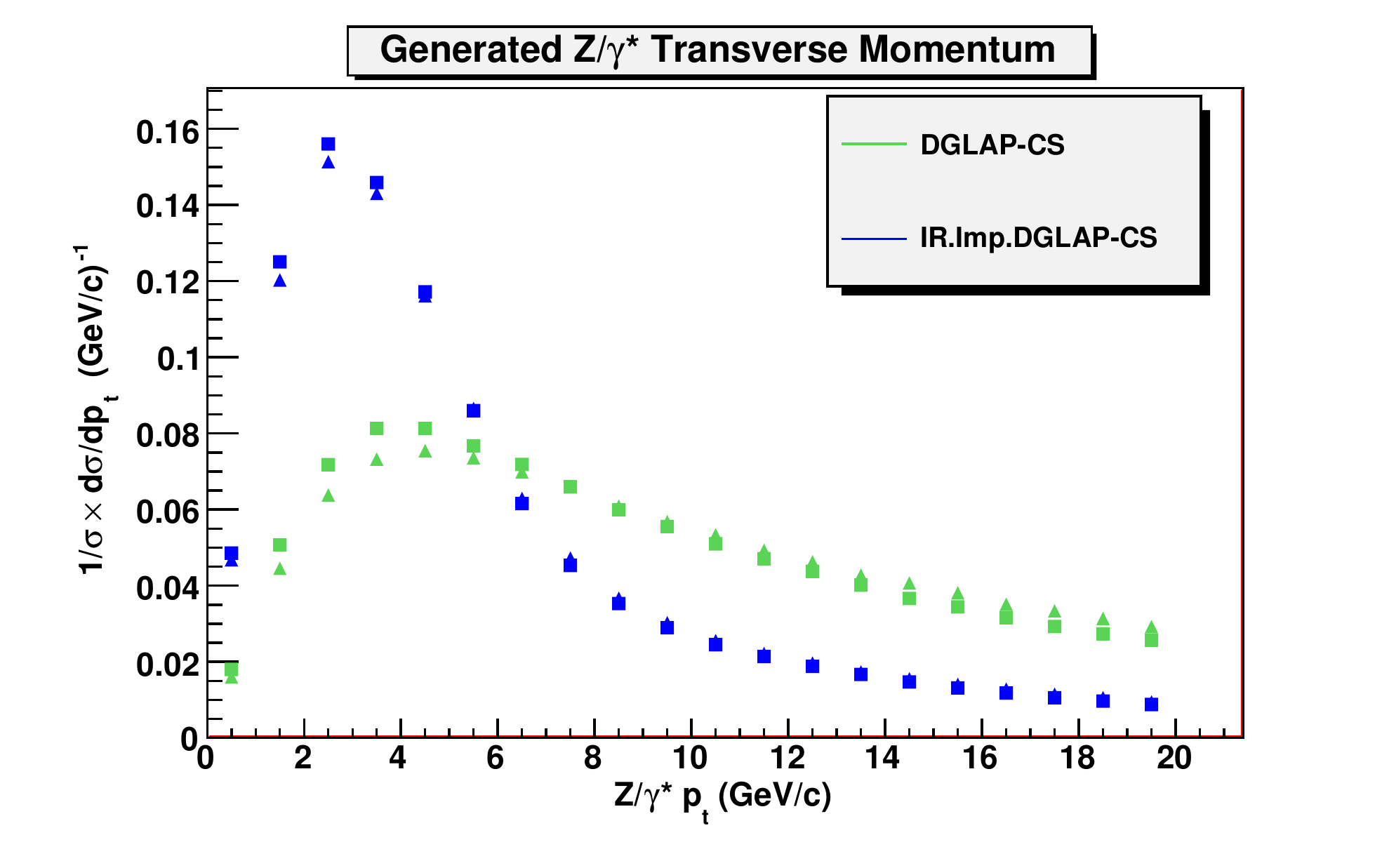}
\end{center}
\caption{\baselineskip=8pt Comparison between IR-improved and unimproved $p_T$ spectra at the LHC as predicted by Pythia8 for single Z/$\gamma^*$ production at cms energies 7 TeV and 13 TeV: blue(green) squares correspond to IR-improved(unimproved) results for 7 TeV cms energy; triangles correspond to the analogous results for 13 TeV cms energy. In black and white print, blue(green) corresponds to dark(light). The results presented here are untuned. 
}
\label{fig1-pyiri}
\end{figure}

While detailed illustrations of the resulting IR-improved phenomenology will 
appear elsewhere~\cite{elswh} as we have noted, here we will 
use the $p_T$ spectrum in single heavy gauge boson production at the LHC 
to illustrate the expected size of the IR-improvement effects in the Pythia8 
environment. Accordingly, we show in Fig.~\ref{fig1-pyiri} the $p_T$ 
spectrum at the LHC for single Z/$\gamma^*$ production when the cms energy is 7 TeV and 13 TeV. We see that the IR improvement has the similar size 
effect at 7 TeV as we have seen in Herwiri1.031~\cite{herwiri3,herwiri4,1412-8717} in the Herwig6.5 environment. Here, we stress that the results in Fig.~\ref{fig1-pyiri} have no intrinsic $p_T$ for the partons in the incoming beams. But, we note that it increases the unimproved Pythia8 prediction in the first bin without the need of ad hoc manipulations as presented in
Ref.~\cite{cteq-13}. We can see from the comparisons between Pythia8(Pythia6) and ATLAS data in Fig. 10(Fig. 8) of Ref.~\cite{atlas-pt} that the increase in the first bin regime is in the right direction to improve the agreement with the data {\it without ad hoc parameter manipulations} and this will be studied in more detail elsewhere~\cite{elswh}. 
These effects must be taken into account in analyzing 
the LHC data in the context of precision QCD for LHC physics, be it 
backgrounds for discoveries or SM tests.\par
To sum up, we have introduced the IR-improved DGLAP-CS theory into the 
showers in Pythia8. The size of the effects are similar to those seen 
in Herwiri1.031 in the Herwig6.5 environment. We encourage experimentalists 
to use this IR-improved version of Pythia8 to explore the interplay of 
IR-improvement with estimation of detector effects, especially when high 
precisions on differentially exclusive spectra are desired. The IR-improved 
version of Pyhtia8 may be obtained 
from the website: http://bflw.web.cern.ch/  .\par
In closing, we
thank Prof. Ignatios Antoniadis and Prof. W. Lerche for the support and kind 
hospitality of the CERN TH Unit while part of this work was completed. We also thank Profs. T. Sjostrand and P. Skands and Dr. J. Christiansen for helpful discussion. \par

\end{document}